\documentclass{elsart}
\usepackage{amsmath,amssymb,amscd,latexsym,cite,verbatim}

\newcommand{\alin}[2]{ \begin{align*} #2 \end{align*} }
\newcommand{\dcite}[1]{\cite{#1}}
\newcommand{\eqn}[2]{ \begin{equation*} #2  \end{equation*} }

\newcommand{\lalign}[2]{ \begin{align} #2 \end{align} }
\newcommand{\leqn}[2]{\begin{equation} #2 \label{#1} \end{equation}}
\newcommand{\leqnarr}[2]{ \begin{eqnarray} #2 \end{eqnarray} }
\newcommand{\lgath}[2]{ \begin{gather} #2 \end{gather} }
\newcommand{\sect}[2]{\section{#2}\label{#1}}
\newcommand{\subsect}[2]{\subsection{#2}\label{#1}}

\newcommand{\DS}{\displaystyle}
\renewcommand{\Re}{\text{\textbf{Re}}}
\renewcommand{\Im}{\text{\textbf{Im}}}
\newcommand{\abs}[1]{\lvert#1\rvert}
\newcommand{\norm}[1]{{\lVert#1\rVert}}
\newcommand{\ihalf}{\ensuremath{\textstyle\frac{i}{2}}}
\newcommand{\ra}{\rightarrow}

\DeclareMathOperator{\atpt}{\hbox{\vrule height3ex depth1.6ex\hskip0.3em}}
\DeclareMathOperator{\diag}{diag}       
\DeclareMathOperator{\spann}{span}

\newcommand{\Ac}{\mathcal{A}}
\newcommand{\Ad}{\mathrm{Ad}}
\newcommand{\Bc}{\mathcal{B}}
\newcommand{\Eb}{\mathbf{E}}
\newcommand{\Eh}{\hat{E}}
\newcommand{\Fh}{\widehat{F}}
\newcommand{\Go}{G_{o}}
\newcommand{\Ko}{K_{o}}
\newcommand{\Lagr}{\mathcal{L}} 
\newcommand{\La}{\mathbf{\Lambda}} 
\newcommand{\Lg}{\Lambda}
\newcommand{\Lm}{\Lambda_{-}} 
\newcommand{\Lmp}{\Lambda_{\mp}}
\newcommand{\Lo}{\Lambda_{0}}
\newcommand{\Lp}{\Lambda_{+}} 
\newcommand{\Lpm}{\Lambda_{\pm}}
\providecommand{\Lra}{\Leftrightarrow}
\newcommand{\Lt}{\widetilde{L}} 
\newcommand{\Mt}{\widetilde{M}} 
\newcommand{\Pbb}{\mathbb{P}}
\newcommand{\Pc}{\check{P}}
\newcommand{\Ph}{\hat{P}}
\newcommand{\Phit}{\widetilde{\Phi}}
\newcommand{\Qt}{\widetilde{Q}}
\providecommand{\Ra}{\Rightarrow}
\newcommand{\Rh}{\hat{R}}
\providecommand{\Rset}{\mathord\mathbb{R}}
\newcommand{\Rt}{\widetilde{R}}
\newcommand{\Sset}{S_\lambda}
\newcommand{\Uh}{\hat{U}}
\newcommand{\Ut}{\widetilde{U}}
\newcommand{\ad}{\mathrm{ad}}
\newcommand{\cGo}{\check{G}_{o}}
\newcommand{\cW}{\mathcal{W}} 

\providecommand{\eb}{\mathbf{e}}
\providecommand{\fb}{\mathbf{f}}
\newcommand{\fforall}{{\;\;\forall\;\;}} 
\newcommand{\g}{\mathfrak{g}} 
\newcommand{\gammah}{\hat{\gamma}}
\newcommand{\gl}{\mathfrak{g}\mathfrak{l}}
\newcommand{\go}{\mathfrak{g}_{o}}
\newcommand{\h}{\mathfrak{h}} 
\newcommand{\hb}{\mathbf{h}}
\newcommand{\kf}{\mathfrak{k}} 
\newcommand{\kk}{\varkappa}
\newcommand{\pih}{\hat{\pi}} 
\newcommand{\prEh}{\pi_{\Eh}}
\newcommand{\prP}{\pi_{\Pc}}
\newcommand{\prPGo}{\pi_{\Pc\!\times\! \Go}}
\newcommand{\prPh}{\pi_{\Ph}}
\newcommand{\prQ}{\pi_{\Qt}}
\newcommand{\psb}{{\bar{\psi}}}
\newcommand{\psh}{{\hat{\psi}}}
\newcommand{\psib}{\bar{\psi}}
\newcommand{\psih}{\hat{\psi}}
\newcommand{\psip}{\psi'}
\newcommand{\rhoi}{\tilde{\rho}}
\newcommand{\rl}{\rho_{\lambda}}
\newcommand{\sL}{\mathfrak{s}\mathfrak{l}}
\newcommand{\su}{\mathfrak{s}\mathfrak{u}}
\newcommand{\xo}{x_{o}}
\newcommand{\yo}{y_{o}}

\begin{document}


\begin{frontmatter}

\title{Spherical symmetry of generalized EYMH fields}
\author{H.P. K{\"u}nzle}
\ead{HP.Kunzle@UAlberta.ca}
\and
\author{Todd A. Oliynyk$^{1}$}\thanks{Present address: School of Mathematical Sciences, Monash University, VIC 3800 Australia}
\ead{todd.oliynyk@sci.monash.edu.au}
\address{Department of Mathematical and Statistical Sciences\\
University of Alberta, Edmonton Canada T6G 2G1}

\begin{abstract}
The possible actions of symmetry groups on generalized Higgs fields coupled to
an Einstein-Yang-Mills field are studied with differential geometrical
techniques involving principal and associated bundles. A classification of
conjugacy classes of these actions and the form of the corresponding invariant
Einstein-Yang-Mills-Higgs (EYMH) fields is obtained and then applied to the
case of static spherically symmetric fields over four dimensional
space-time. The representations of the gauge group for which spherically
symmetric Higgs fields exist are identified and the set of all field equations
for the independent functions that describe these fields is analyzed and the
corresponding ordinary system of differential equations is derived and shown
to be consistent.
\end{abstract}

\end{frontmatter}

\sect{chintro}{Introduction}

It has long been realized \dcite{k4006} that Yang-Mills
potentials correspond to the local functions needed to describe a
connection on a principal bundle $P$ over space-time $M$ whose
structure group $\Go$ is the physical gauge group. A local gauge
transformation is then represented by a change to another local
section of $P$.

Similarly, while the standard Higgs field is a scalar function on
space-time with values in the Lie algebra $\go$ of $\Go$ and
transforms under the adjoint transformation in $\go$, it is
easily generalized to have values in a vector space (or manifold)
on which the gauge group acts. In fact, a \emph{generalized Higgs
field} is best defined as a section of a bundle $E$ associated to
$P$ since that already incorporates the relation to the gauge
group and the gauge changes \dcite{k0676,k1030,k0832,k5353}.

Over the last one or two decades much work has been done exploring special
classical solutions of Yang-Mills gauge fields in interaction with the
gravitational field, the so-called Einstein-Yang-Mills (EYM) fields.  This
work starting with numerical regular and black hole solutions for the $SU(2)$
gauge group \dcite{k4752,hka25,k5062,k4971} and followed soon by rigorous
existence proofs \dcite{k5061,k5201,k5428} has dealt mainly with the
spherically symmetric static case although some very special rotationally
symmetric stationary solutions have also been numerically constructed.  In
most papers the gauge group was chosen to be the simplest nonabelian one,
namely $SU(2)$, but some numerical studies have been done for $SU(n)$ with
$n\ge3$. The inclusion of gravity unfortunately makes some of the techniques
used in the Yang-Mills-Higgs theories on Minkowski space less effective like,
for example, the Bogomol'nyi equations which have lead to many rigorous
results for arbitrary compact gauge groups (cf. \dcite{k0764}. For a survey of
EYM solutions see \dcite{k6373}1).

We have been particularly interested in studying the general geometric,
analytic and algebraic problems that arise when the gauge group is an
arbitrary compact semisimple Lie group and the symmetry group acts on the
principal bundle by arbitrary automorphisms as long as they project onto the
`normal' action of $SO(3)$ by isometries on a static space-time manifold.  In
\dcite{k5281,eymg} it was shown how the the conjugacy classes of these group
actions correspond to Dynkin's \dcite{k4779} classification of
$\mathfrak{sl}_2\mathbb{C}$ subalgebras of semisimple Lie algebras. We then
derived and analyzed to some extent the resulting system of ordinary
differential equations \dcite{eymg,eymgg,eymgl}.  It turns out that the
analysis of these equations even in this static spherically symmetric case
poses already many interesting problems. About the set of global solutions
satisfying appropriate physical boundary conditions still very little is known
for gauge groups other than $SU(2)$. The possible actions of $SU(2)$ by
automorphisms on principal bundles come in two main groups which we called
regular and irregular. For regular actions the static field equations allow a
simple gauge choice such that with suitable boundary conditions a nonlinear
and singular boundary value problem for a number of functions of the radial
variable $r$ results which is numerically difficult to solve but feasible
(see, for example, \dcite{k5749,k6174}). In the irregular case there appears
to be no simple gauge choice to eliminate some dependent variables and the
boundary value problem becomes degenerate and thus numerically quite unstable.

The purpose of this paper is two-fold. On the one hand we analyze carefully
the possible actions of a symmetry group on both the principal and associated
bundles and extend the classification of the conjugacy classes of action by
automorphisms on principal bundles by Brodbeck \dcite{k6015} to those on
associated bundles. This construction works for quite general symmetry group
actions, structure groups and representations of the structure group on vector
bundles subject to only very mild restrictions on the orbit structure. The
main result is Theorem 1 which also allows us to make a fairly natural gauge
choice in the general and in particular also the spherically symmetric case.

On the other hand we use this general result to classify the possible static
spherically symmetric field equations of a general Einstein-Yang-Mills-Higgs
(EYMH) system for arbitrary compact semisimple gauge groups, arbitrary
symmetry group actions, and arbitrary representations defining the associated
bundle whose sections are the generalized Higgs fields. This classification
also applies, of course, to the Yang-Mills-Higgs theory where some early work
\dcite{k0943} was done before the study of the EYM equations started. When
generalized Higgs fields are included many results of the representation
theory of compact Lie groups can be used.

We work with theories for which the Lagrange density is of the form $\Lagr
\sqrt{\abs{g}}d^4x$ with \leqn{lagr}{ \Lagr = \kk R - 2\Lambda -
  k(F_{\alpha\beta},F^{\alpha\beta}) - h(D_{\alpha}\Phi,D^{\alpha}\Phi) -
  \cW\bigl(\,h(\Phi,\Phi)\,\bigr), } where $R$ is the scalar curvature of the
metric $g_{\alpha\beta}dx^\alpha dx^\beta$, $k$ is an ad-invariant positive
definite inner product on the Lie algebra $\g_{o}$ of the gauge group $\Go$,
and $h$ a (Hermitian) inner product on a (in general complex) vector space
$V$, invariant under the action $\rho:\Go\ra GL(V)$ and $\cW$ is a scalar
function of its argument serving as a potential.  $D_{\alpha}\Phi$ denotes the
gauge covariant derivative of the Higgs field $\Phi$ which depends on the
metric, the gauge potential and $\rho$. (Also, $\kk=c^4/(8\pi G)$ with $G$
being Newton's constant, $\Lambda$ is the cosmological constant while coupling
constants for the Yang-Mills and Higgs fields can be absorbed into the
definitions of the inner products $k$ and $h$.)

Part of our assumptions is that both the gauge field and the Higgs field are
invariant under the appropriate actions of the symmetry group on the principal
and the associated bundle, respectively. We then find that nontrivial
spherically symmetric Higgs fields may not exist for certain representations
$\rho$ like a 2-dimensional irreducible spinor representation, for example.
This does not exclude the possibility of such a (noninvariant) Higgs field
contributing to a spherically symmetric stress-energy tensor, however, and
thus being compatible with a spherically symmetric space-time metric. But we
are not aware of any reasonable definition of symmetry in which such Higgs
fields themselves could be regarded as spherically symmetric.

The paper is organized as follows. In sections \ref{assocb} and \ref{sym} we
recall the definition of automorphisms and automorphism groups of associated
bundles and establish some notation. The general classification of invariant
gauge and Higgs fields under any symmetry group action is obtained in section
\ref{symC}. In section \ref{eqs} we specialize to the symmetry group $K=SU(2)$
and derive all field equations. We then verify that a consistent set of first
and second order ordinary differential equations in the radial variable is
obtained, subject to a set of constraint equations that need be satisfied only
at one regular point and are then `conserved'.  This result is true whether or
not the action of $K$ is regular or not, but as in the pure EYM case the gauge
choice is simple only in the regular case.  In the final subsection we use
results from \dcite{eymg,eymgg} to cast the field equations into a fairly
explicit form from which a numerical algorithm could be derived. We do not, in
this paper, analyze what kind of boundary conditions are implied by reguarity
assumptions on the solution at singular points like the center ($r=0$) or at a
black hole horizon. In the general case (arbitrary compact gauge group and
arbitrary representation) this is likely to lead to a considerable number of
nontrivial algebraic problems as one can guess from the experience with the
EYM fields. It is, however, a necessary first step for a numerical exploration
of the solution set.

\sect{assocb}{Associated bundles and their automorphisms}

Let $P=(P,\pi,M,\Go,R)$ be a principal bundle over a manifold $M$ with
projection $\pi$, structure group $\Go$ and right action $R$ of $\Go$ on $P$,
$R:P\times \Go\ra P:(p,g)\mapsto R_{g}p$.

Given another manifold $V$ and a left action $\rho:\Go\times V\ra V$ the
\emph{associated bundle} $E=(E,P,\pi_{E},M,\Go,R,\rho,V)$ is defined as the
set of equivalence classes $[p,v]$ of elements $(p,v)\in P\times V$ with
respect to the relation
\eqn{equiv}{
(p',v')\sim(p,v) \Lra p'=R_{g}p \text{\ and\ } v'=\rho_{g^{-1}}v 
\text{\ for some\ }g\in \Go.
}
We denote the fibers of $P$ and $E$ over $x\in M$ by $P_{x}=\pi^{-1}(x)$ and
by $E_{x}=\pi_{E}^{-1}(x)$, respectively.

We will, in general, assume that the action $\rho$ is \emph{effective}, i.e.
that $\rho_{g}v=v\;\forall\;v\in V\Rightarrow g=e$, the identity of $\Go$.  It
then follows that $p,q \in P_{x},\;\; [p,v] = [q,v]\;\forall\;v\in V \; \Ra \;
p=q$.
Let $\pih:P\times V\ra E:(p,v)\mapsto [p,v]$ denote the canonical
projection. Then
$\pih_{p}:V\ra\pi_{E}^{-1}(\pi(p)):v\mapsto [p,v]$
is an isomorphism of $V$ onto $E_{\pi(p)}$ (a diffeomorphism in
general, a vector space isomorphism if $V$ is a vector space). We
note that $\pih_{p}^{-1}\left([p,v]\right)=v$ and also
$\pih_{R_{g}p} = \pih_{p}\circ \rho_{g}$.

It is well known (e.g. \dcite{k5353}) that there is a one-to-one
correspondence between the set $C\pi_{E}$ of sections $\Phi$ of
$E$ and equivariant maps $\tilde{\Phi}:P\ra V$ (i.e.  maps
satisfying $\Phit\circ
R_{g}=\rho_{g^{-1}}\circ\tilde{\Phi}\;\forall\;g\in \Go$) given
by $\Phi \mapsto \Phit$ with $\Phit(p) =
\pih_{p}^{-1}\circ\Phi\circ \pi(p)$ and $\tilde{\Phi} \mapsto
\Phi$ with $\Phi(x) = \pih_{p}\circ\tilde{\Phi}(p) =
[p,\tilde{\Phi}(p)]$ for any $p\in P_{x}$ and $x\in M$.

An \emph{automorphism of $P$} is a diffeomorphism $\psi$ of $P$
onto itself such that
$\pi\circ\psi =\psb \circ \pi \quad\text{and}\quad \psi\circ
R_{g}=R_{g}\circ\psi \;\forall\; g\in \Go$
where $\psb$ is an induced diffeomorphism of $M$ onto itself.
An \emph{automorphism of $E$} is a bundle isomorphism
$(\chi,\psb)$ of $E$, i.e. satisfying $\pi_{E}\circ\chi =\psb
\circ \pi_{E}$, and inducing an isomorphism of $E_{x}$ onto
$E_{\psb(x)}$ for any $x\in M$ that is of the form
$\chi=\pih_{q}\circ\pih_{p}^{-1}: E_{x} \ra E_{\psb(x)} : [p,v]
\mapsto [q,v]$
for a certain $p\in P_{x}:=\pi^{-1}(x)$ and $q\in P_{\psb(x)}$
(cf.\dcite{k0929},{p.55}). 

Given an automorphism $\psi$ of $P$ there is, however, a natural way to induce
a related automorphism $\psi^{E}$ of an associated bundle $E$, namely by
choosing $q=\psi(p)$ so that
\leqn{ppsp}{
\psi^{E} = \pih_{\psi(p)}\circ\pih_{p}^{-1} : E_{x} \ra E_{\bar{\psi}(x)} : [p,v] \mapsto
 [\psi(p),v] \;\;\text{for any $p\in P_{x}$}.
}
In this case 
\emph{a section $\Phi\in C\pi_{E}$
is invariant under an automorphism $\psi_{E}$ of $E$,
i.e. satisfies}
$\Phi \circ \bar{\psi} = \psi^{E} \circ \Phi$

iff the corresponding equivariant map $\tilde{\Phi}:P\ra V$ is
invariant in the sense of satisfying
$\tilde{\Phi}\circ\psi = \tilde{\Phi}.$

(See \dcite{k1641}, for example.)
Every automorphism of $E$ is induced by one of $P$ in this
way, provided that the action $\rho$ defining $E$ is effective. 

We now describe both $\psi$ and $\psi^{E}$ with respect to a
local trivialization $U\times \Go$ of $P$ where $U$ is an open
set of $M$.  Given a local section $\sigma:U\ra P$ define a local
trivialization of $P$ by
\leqn{triv}{
\tau:U\times \Go\ra \pi^{-1}(U):(x,g)\mapsto R_{g}\sigma(x).
}
and let
\leqn{trivE}{
\tau_{E}:U\times V\ra \pi_{E}^{-1}(U):(x,v)\mapsto [\sigma(x),v]
}
be the associated local trivialization of $E$. With respect
to this trivilization a section $\Phi$ of $E$ can always
be written
as 
\leqn{PhiT}{
\Phi(x) = \tau_{E}(x,\phi(x)) = [\sigma(x),\phi(x)] \quad x\in U
}
for some map $\phi : U \rightarrow V$.

The corresponding equivariant map $\tilde{\Phi}$ then satisfies
$\tilde{\Phi} \circ \tau(x,g) = \rho_{g^{-1}} \phi(x)$
and under a gauge change a Higgs field $\Phi$ transforms like
$\phi(x) \mapsto \check{\phi}(x) = \rho_{\gamma(x)^{-1}}\phi(x)$.

In such a local chart we can describe the automorphism $\psi$ in
the form
$\psi\bigl(\sigma(x)\bigr) = R_{\psh(x)}\sigma\bigl(\psb(x)\bigr)$
where $\psh: U\ra \Go$. Then
$\psi\big(\tau(x,g)\bigr) = \tau\bigl(\psb(x),\psh(x)g\bigr)$.
Similarly, the map 
$\psi^{E}\bigl([\sigma(x),v]\bigr) := [
R_{\psi^{E}(x)}\sigma(\psb(x)), v ] = [ \sigma(\psb(x)),
\rho_{\psi^{E}(x)}v ]$
is locally of the form
$\psi^{E}\circ \tau_{E}(x,v) 
\tau_{E}\bigl(\psb(x),\rho_{\psh(x)}v\bigr)$.
Under a local gauge transformation
$\sigma_{2}(x)=R_{\gamma(x)}\sigma_{1}(x)$ the functions
$\hat{\psi}_{1}$ changes into $\hat{\psi}_2 =
\gamma\bigl(\bar{\psi_{1}}(x)\bigr)^{-1}\hat{\psi}_{1}(x)
\gamma(x)$.

\sect{sym}{Symmetry group acting on $P$ and $E$}

We will be interested in groups of automorphisms of $P$ and of
$E$ that cover the same diffeomorphisms of $M$ and want to
explore just how many independent choices can be made to describe
such actions completely.  For classical relativistic field
theories we would expect to have an isometry group of a
Lorentzian space-time manifold $M$ and, if there are gauge fields
and Higgs fields present, we would expect this group to lift to
act by automorphisms on the bundles $P$ and $E$ so that all
physical fields are invariant under this symmetry group
action. (This is to some extent even implied if Einstein's
equations hold because then, if the metric is invariant, the
whole stress-energy tensor will have to be invariant too which
imposes strong constraints on the gauge and Higgs fields although
it does not imply that they are invariant under the symmetries.)

For the remainder of this article we will assume that that the
associated bundle $E=P\times_{\rho} V$ is a \emph{vector bundle}.
That is $V$ will be taken to be a finite dimensional vector space
and $\rho : \Go \rightarrow \text{GL}(V)$ a \emph{linear
representation} of $\Go$ on $V$. We will also assume that there
is a positive definite Hermitian inner product $h:V \times V
\rightarrow \Rset$ on $V$ that is invariant under the action of
$\Go$.  In other words $h(\rho_{g}v,\rho_{g}w) = h(v,w)$ for all
$g\in \Go$ and $v,w\in V$. We note that if $\Go$ is compact, then
there will always exist such an inner-product.

We call a principal bundle $P=P(M,\Go)$ on which a Lie group
$K$ acts effectively on the left 
\leqn{symauto1}{
\psi:  K\times P \longrightarrow P\: :\:
(a,p)\longmapsto \psi_{a}p
}
by principal bundle automorphisms
a \emph{$K$-symmetric principal bundle}.
Let
$\bar{\psi} : K\times M \longrightarrow M \: :\: 
(a,x) \longmapsto \bar{\psi}_{a}x$
denote the left action of $K$ induced on $M$
via projections of $\psi_{a}$.
As discussed in the previous section, the action \eqref{symauto1}
induces a \emph{natural} left action of $K$ on $E$ by
bundle automorphisms which is given by
\leqn{symauto2}{ 
\psi^{E}:  K\times E \longrightarrow E\: :\: 
(a,[p,v]) \mapsto [\psi_{a}(p),v]. 
}
and in local coordinates by
\leqn{symloc}{
  \psi_{a}(x,g) = (\psb_{a}x,\psh(a,x)g)\quad\text{and}\quad
  \psi^{E}_{a}(x,v) = (\psb_{a}x,\rho_{\psh(a,x)}v). 
} 
The symmetry group action is therefore determined (in a given gauge) by
two maps $\psb: K\times U\ra U$ and $\psh: K \times U\ra \Go$. The fact that $K$
acts on the left implies
$\psb(ab,x)= \psb(a,\psb(b,x))$, $\psh(e_{K},x)   =  e_{\Go}$, 
and $\psh(ab,x)= \psh(a,\psb_{b}x)\psh(b,x)$ for
all $a,b\in K$ and $x\in U$.

It is easily seen that $\psi_{E}$ induces a right action $\psi_{E}^{*}$ on the set
$C\pi_{E}$ of sections of $E$ by
\leqn{actsec}{
 \psi^{E}_{a}{}^{*}(\Phi) := \psi^{E}_{a^{-1}}
\circ\Phi\circ\psb_{a}\quad\forall\;a\in K.
}
Therefore a  section $\Phi$ is 
called \emph{invariant under the action of $K$} if
\eqn{Kinvar}{
\psi^{E}_{a}{}^{*}(\Phi)=\Phi \;\forall\;a\in K.
}

The invariant Hermitian inner-product $h$ can be
used to induce a Hermitian inner product on the vector
bundle $E$. If $\sigma : U\subset M \rightarrow P$ is a local
section and $\Phi$, $\Psi$ $\in C\pi_{E}$ are two sections
with local representatives $\Phi^{\sigma} : U
\rightarrow V$ and $\Psi^{\sigma} : U \rightarrow V$,
respectively, so that
\eqn{locrep}{
\Phi(x) = [\sigma(x),\Phi^{\sigma}(x)]
\quad \text{and} \quad
\Psi(x) = [\sigma(x),\Phi^{\sigma}(x)] \quad \text{for all
$x \in U$, }
}
then the Hermitian metric $h$ on $E$
is defined by the formula
\leqn{Emetric}{
h(\Phi,\Phi) := h(\Phi^{\sigma},\Psi^{\sigma}) \quad \text{for all
$x\in U$.}
}
The $\Go$-invariance of $h$ guarantees that this local
formula defines a global Hermitian metric. 

\sect{symC}{Classifying invariant Higgs fields}

As in the previous section we assume that $P=P(M,\Go)$ is a
$K$-symmetric bundle and that $K$ acts on the vector bundle
$E=P\times_{\rho} V$ according to the natural action
\eqref{symauto2}. Also, for the remainder of this article we will
assume that the symmetry group $K$ is \emph{compact}. Once we
know this, then we know that there exists an open dense subset
$U\subset M$ such that $U$ is, at least locally, regularly
foliated by orbits of $K$ under the action $\psb$ on M. Fixing a
point $x_{o} \in U$ and letting $\Ko$ be the isotropy group of
$x_{o}$, we then have that locally $U \approx U/K \times K/\Ko$.
This shows that we can, with a minor loss of generality, assume
that $M = \Mt\times K/\Ko$ with the $\psb$ action given by
\eqn{psiA}{ \psb : K\times (\Mt\times K/\Ko) \longrightarrow
\Mt\times K/\Ko \: :\: (a,(x,k\Ko)) \longmapsto (x,ak\Ko) \, .  }

In \dcite{k6015} it is established that the $K$-symmetric
principal bundles over base manifolds of the form $\Mt\times
K/\Ko$ can be classified by a homomorphism $\lambda : \Ko
\rightarrow \Go$ and a principal bundle $\Qt$ over $\Mt$ with
structure group $Z := \text{Cent}(\lambda(\Ko)) \subset \Go$.
The classifying bundle $\Qt$ is constructed as follows.  Let
$P\bigl|_{\Mt}$ be the portion of $P$ over the submanifold $\Mt
\cong \Mt \times \{e\Ko\}$ (i.e. $P\bigl|_{\Mt}:=\{p\in
P\,:\,\pi(p)\in\Mt\times{e\Ko}\}$).  Then $\Mt\cong
\Mt\times\{e\Ko\}$ is a fixed point set for the action $\psb$ and
hence each fiber of $P\bigl|_{\Mt}$ is mapped onto itself by the
action of $\Ko$ on $P$. This induces a map \eqn{mumapA}{ \mu :
P\bigl|_{\Mt}\times \Ko \longrightarrow \Go \: : \: (p,h)
\longmapsto \mu_{p}(h) } where $\mu_{p}(h)$ is the unique element
of $\Go$ satisfying $\psi_{h}(p) = R_{\mu_{p}h}(p)$.  For each
$p\in P\bigl|_{\Mt}$, $\mu_{p} : \Ko \rightarrow \Go$ defines a
group homomorphism. Moreover, if $p,q\in P\bigl|_{\Mt}$ are in
the same fiber then the homomorphisms $\mu_{p}$ and $\mu_{q}$
belong to the same conjugacy class.  Next, fix $p_{o} \in
P\bigl|_{\Mt}$ and let $\lambda := \mu_{p_{o}}$.  Then $\Qt$ is
defined by \leqn{Qbundle}{ \Qt(\Mt,Z) := \{ p \in
P\bigl|_{\Mt}\,:\, \mu_{p}= \lambda \} } and it can be shown that
$\Qt$ is a principal bundle over $\Mt$ with structure group $Z$.
Thus each $K$-symmetric principal bundle $P$ with base manifold
$M = \Mt \times K/\Ko$ determines a $Z$-bundle $\Qt$ over $\Mt$
and a homomorphism $\lambda : \Ko \rightarrow \Go$. Conversely,
given $(\Qt,\lambda)$ it is possible to construct a bundle
isomorphic to $P$. We describe the construction below because it
produces a principal bundle $\Ph$ isomorphic to $P$ on which the
$K$ action is made as simple as possible.  This makes it easy to
identify the $K$-invariant Higgs fields.

Let $\check{P} := \Qt \times K$ be the product bundle with base
$\Mt \times K/\Ko$ and gauge group $\cGo:= Z\times \Ko$ which
acts on $\check{P}$ via \eqn{Gpright}{ R_{(z,h)}(q,k) =
(q,k)\cdot (z,h) := (R_{z}(q),kh) \quad \text{for all $(z,h)\in
Z\times \Ko$.}  } The projection $\prP : \check{P} \rightarrow
\Mt\times K/\Ko$ is given by $\prP(q,k) = (\prQ(q),k\Ko)$ where
$\prQ : \Qt \rightarrow \Mt$ is the principal bundle projection
map.  Clearly, \eqn{Kact}{ \psip: K \times \check{P} \ra
\check{P} \,:\, (k_{1},(q,k)) \mapsto (q,k_{1}k) } is a left
action of $K$ on $\check{P}$ by bundle automorphisms.

We let $\rl$ be the homomorphism defined by $\rl :
\cGo\rightarrow \Go \: : \: (z,h) \mapsto z\lambda(h)$, and let
$\cGo$ act on $\Go$ via $\cGo\times \Go \rightarrow \Go \: : \:
(g',g) \mapsto \rl(g')g$.  This allows us to define the
associated bundle $\Ph := \check{P}\times_{\rl} \Go$.  It can be
verified that $\Ph$ is a principal bundle with base $\Mt \times
K/\Ko$ and structure group $\Go$ with the right action of $\Go$
given by \eqn{Phright}{ \Rh_{g_{1}}([(q,k),g]) = [(q,k),g]\cdot
g_{1} := [(q,k), gg_{1}] \qquad g_{1}\in \Go \, .  } The bundle
projection map $\prPh : \Ph \rightarrow \Mt\times K/\Ko$ is given
by $\prPh([(q,k),g]) := \prP(q,k)$.  As indicated above, the
importance of $\Ph$ is that it is isomorphic to $P$ with the
isomorphism defined by $\Ph \rightarrow P \: :\: [(q,k),g]
\mapsto \psi_{k} R_{g}q$.  This defines a $K$ and $\Go$
equivariant bundle isomorphism that induces the identity on the
common base $\Mt\times K/\Ko$ of the two bundles $\Ph$ and
$P$. We also note that left action of $K$ on $\Ph$ is given
simply by \leqn{phiact}{ \psih\,:\, K \times \Ph \longrightarrow
\Ph \,:\, \bigl(k_{1},[(q,k),g]\bigr) \longmapsto [(q,k_{1}k),g]
\, .  }

Now that the bundle $\Ph$ has been defined, we can use it
to classify the invariant Higgs fields.
Consider the associated vector bundle
\eqn{Ebundle}{
\Eh := \Ph \times_{\rho} V \,
}
with projection $\prEh : \Eh \ra \Mt \times K/M$. The points
of $\Eh$ are the equivalence classes $[[(q,k),g],v]$ where $[(q,k),g]$ 
is a point in $\Ph$ and $v$ is a vector in $V$. We note
that the natural left $K$-action $\psi^{\Eh} : K\times \Eh \ra \Eh$ 
is given by
\eqn{Eleft}{
\psi^{\Eh} : K\times \Eh \longrightarrow \Eh \: : \: (k_{1},[[(q,k),g],v]) \longmapsto 
\psi^{\Eh}_{k_{1}}([[(q,k),g],v]) := [[(q,k_{1}k),g],v] \, .
}
This follows from \eqref{phiact} and \eqref{symauto2}.  Let $\Phi : \Mt \times
K/\Ko \rightarrow \Eh$ be a section of $\Eh$ that is $K$-invariant in the sense
of \eqref{actsec}. In other words, $\Phi$ is a $K$-invariant Higgs field. From
section \ref{assocb}, we know that $\Phi$ is equivalent to a $\Go$-equivariant
map $\Phit : \Ph \rightarrow V$ which satisfies $\Phit\circ \psh_{k} = \Phit
\quad \text{for all $k\in K$}$.  Letting $\prPGo$ denote the projection
$\prPGo : \check{P}\times \Go \rightarrow \Ph \: : \: (p,g) \mapsto [p,g]$, we
then find that the diagram
\leqn{bmap}{
\begin{CD}
\check{P}\times \Go @> \prPGo >> \Ph \\
@V\text{pr}_{1}VV  @VV\prPh V\\
\check{P} @>>\prP > \Mt\times K/\Ko
\end{CD}
}
commutes.  We claim that this defines a $K$-equivariant principal bundle
homomorphism between $\Ph$ and $P$. To see this note that
\eqn{bmap1}{
\prPGo\bigl(k_{1}\cdot((q,k),g)\bigr) = \prPGo\bigl(((q,k_{1}k),g)\bigr)
= [(q,k_{1}k),g] = k_{1}\cdot [(q,k),g]
}
which shows that $\prPGo\circ \psip_{k} = \psih_{k}\circ \prPGo$ for all $k\in
K$, where $\psip: \bigl(k_{1},((q,k),g)\bigr) \mapsto ((q,k_{1}k),g)$ is the
natural left action of $K$ on $\check{P}\times \Go$.  Also
\eqn{bmap3}{
\prP(k_{1}\cdot(q,k)) = \prP((q,k_{1}k)) = (\prQ(q),k_{1}k\Ko) = k_{1} \cdot(\prQ(q),k\Ko)
= k_{1} \cdot \prP(q,k)
}
which shows that $\prP\circ\psi_{k} = \psb_{k}\circ \prP$ for all $k\in K$.
This establishes the $K$-equivariance of the bundle map. We also note that
since $\prPGo\bigl(((q,k),g)\cdot g_{1}) = [((q,k),gg_{1}] = [(q,k),g]\cdot
g_{1}R\,$, it follows that $\prPGo \circ \check{R}_{g} = \check{R}_{g} \circ
\prPGo$ for all $g \in \Go$, where we are using
$\check{R}_{g_{1}}((q,k),g) := ((q,k),gg_{1})$ to denote the right action of
$\Go$ on the bundle $\check{P}\times \Go$.  This shows that the bundle map
\eqref{bmap} induces the identity homomorphism on $\Go$.

From the commutative diagram
\eqn{bmap6}{
\begin{CD}
\check{P}\times \Go @> \prPGo >> \Ph @>\Phit >> V\\
@V\text{pr}_{1}VV  @VV\prPh V\\
\check{P} @>>\prP > \Mt\times K/\Ko
\end{CD}
}
and the fact that $\prPGo$ is surjective, it is clear that the equivariant map
corresponding to the Higgs field $\hat{\Phi}$ is completely determined by the
map
\eqn{Psit}{
\check{\Phi} := \hat{\Phi} \circ \prPGo : \check{P}\times \Go \longrightarrow
V \, .
}
Since $\prPGo$ is both $\Go$ and $K$ equivariant,
it follows that $\check{\Phi}$ is $K$ and $\Go$ equivariant. That
is $\check{\Phi}$ satisfies
\leqn{Psiequi}{
\check{R}_{g} \circ \check{\Phi} = \rho(g^{-1})\circ \check{\Phi} \quad \text{
for all $g\in \Go$} \,
}
and 
\leqn{PsisymA}{
\check{\Phi}\circ \psip_{k} =  \check{\Phi}  \quad \text{for all $k\in K$.}
} 
The map $\check{\Phi}$ possesses an additional invariance coming from the
construction of the bundle $\Ph$. Letting $\phi_{g}$ denote the action
\eqn{Gpact}{
\phi_{g'} : \check{P}\times \Go \longrightarrow \check{P}\times \Go \: : \:
((q,k),g) \longmapsto (R_{g'}((q,k)),\rho_{\lambda}({g'}^{-1})g) \quad
\text{for\ }g'\in \cGo
}
it follows from the definition of $\Ph$ as the associated bundle that $\prPGo
\circ \phi_{g'} = \prPGo$ for all $g'\in \cGo$.  Consequently $\check{\Phi}$
satisfies
\leqn{PsisymC}{
\check{\Phi} \circ \phi_{g'} = \check{\Phi} \quad \text{ for all $g'\in \cGo$.}
}
From \eqref{Psiequi} we have that
$\check{\Phi}((q,k),g) = \rho_{g^{-1}}\check{\Phi}((q,k),e)$
while \eqref{PsisymA} shows that
$\check{\Phi}((q,k),g) = \check{\Phi}((q,e),g)$.
Combining these two results yields
\leqn{relC}{
\check{\Phi}((q,k),g) = \rho_{g^{-1}}\check{\Phi}((q,e),e) \, .
}
We also have from \eqref{PsisymC} that
\leqn{relD}{
\check{\Phi}((q,k),g)) =  \rho_{g^{-1}}
\check{\Phi}((q\cdot z,kh),z^{-1}\lambda{h^{-1}}g)\, .
}
Equations \eqref{relC} and \eqref{relD} imply that
\leqn{relE}{
\check{\Phi}((q\cdot z,e),e) =
\rho_{z^{-1}}\rho_{\lambda(h^{-1})}\check{\Phi}((q,e),e)
}
for all $q\in\Qt$ and $z\in Z$, $h\in \Ko$. Defining the map
\leqn{Lmap}{
\Lt : \Qt \longrightarrow V \: :\: q \longmapsto \Lt(q) := 
\check{\Phi}((q,e),e)\, ,
}
equation \eqref{relC} shows that $\Lt$ and the action $\rho$ can be used to
completely determine the invariant Higgs field via the relationship
$\Phit([(q,k),g]) = \rho_{g^{-1}}\Lt(q)$ for all $q\in \Qt$, $k\in K$, and
$g\in \Go$.  This can also be written as
\leqn{HiggsLB}{
\Phi\circ \prP(q,k) = [[(q,k),g],\rho_{g^{-1}}\Lt(q)]
\quad \text{for all
$q\in \Qt$, $k\in K$, and $g\in \Go$.}
}

The map $\Lt$ is not arbitrary but is in fact a section of the associate
bundle $\Qt\times_{(\rho,Z)}V$. To see this set $h=e$ in \eqref{relE} to get
$\Lt\circ R_{z} = \rho_{z^{-1}}\circ \Lt$ for all $z\in Z$.  This shows that
$\Lt :\Qt \rightarrow V$ is an equivariant map and hence by the discussion in
section \ref{assocb}, uniquely determines a section of the vector bundle
$\Qt\times_{(\rho,Z)}V$.  Setting $z=e$ in \eqref{relE} shows that $\Lt$ must
satisfy also the additional condition
\leqn{LmapsymA}{
\rho_{\lambda(h)}\Lt(q) = \Lt(q) \quad \text{for all $q\in \Qt$
and all $h\in \Ko$.}
} 

We summarize the above results in the following theorem:
\begin{thm} \label{HCthm} 
Let $P(M,\Go)$ be a principal bundle and suppose that
\begin{itemize}
\item[(i)]$K$ is a compact Lie group that acts on $P(M,\Go)$ on the left by
principal bundle automorphisms,
\item[(ii)] the base space $M$ is diffeomorphic to $\Mt\times K/\Ko$ where
$\Mt$ is a smooth manifold, $\Ko$ is the isotropy subgroup of $K$ of any point
$x_{o} \in M$, and the induced action of $K$ on $M$ is given by
$(k_{1},(x,k\Ko)) \rightarrow (x,k_{1}k\Ko)$,
\item[(iii)] the $K$-symmetric bundle $P(M,\Go)$ is classified by a
homomorphism $\lambda : \Ko \rightarrow \Go$ and a principal bundle
$\Qt(\Mt,Z)$ ($Z:=\text{\emph{Cent}}(\lambda(\Ko))\subset \Go$) in the sense
of Brodbeck (see \eqref{Qbundle} and \dcite{k6015}),
\item[(iv)] $V$ is a vector space, $\rho$ is a linear representation of $\Go$
on $V$, and $E:=P\times_{\rho}V$ is the associated vector bundle.
\end{itemize}
Then the set of $K$-invariant sections of the vector bundle $E$ is in
one-to-one correspondence with the set of maps $\Lt : \Qt \rightarrow V$
satisfying
\eqn{HCthm1}{
\Lt \circ R_{z} = \rho_{z^{-1}} \circ \Lt \quad \text{and} \quad 
\rho_{\lambda(h)}\circ \Lt = \Lt
}
for all $z\in Z$ and $h\in \Ko$.
\end{thm}

In order to derive the EYMH equations we need to have explicit
local formulae for the invariant fields.  This means fixing a
gauge.  To fix the gauge, let $\tilde{\sigma} : \Ut \subset \Mt
\rightarrow \Qt$ and $\hat{\sigma} : \Uh \subset K/\Ko
\rightarrow K$ be two local sections.  Then, if $\iota_{e} :
\check{P} \rightarrow \check{P}\times \Go\: : \: p \mapsto
(p,e)$, the two local sections $\tilde{\sigma}$ and
$\hat{\sigma}$ can be used to define local section (i.e. a gauge)
of $\Ph$ by
\leqn{lf3}{
\sigma : \Ut\times \Uh \subset \Mt\times K/\Ko \rightarrow \Ph
\quad \sigma := \prPGo\circ \iota_{e} \circ \tilde{\sigma}\times
\hat{\sigma} \, .
}
In \dcite{k6015} it is shown that the the set of $K$-invariant
connection forms $\omega$ on $P$ is in one-to-one correspondence
with the the set of pairs $(\tilde{\omega},\tilde{\Lambda})$
where $\tilde{\omega}$ is a connection form on $\Qt$ and
$\tilde{\Lambda}$ is a map $\tilde{\Lambda} : \Qt\longrightarrow
\gl(\kf,\g)$ that satisfies
\leqn{CCthm2}{
\tilde{\Lambda}(q)\circ \Ad_{h} = \Ad_{\lambda(h)}\tilde{\Lambda}(q)
\quad \text{and} \quad \tilde{\Lambda}(q)\cdot \xi =
\lambda'(\xi) \quad (\lambda':= T_{e}\lambda) 
}
for all $q\in \Qt$, $h\in \Ko$, and $\xi \in \kf_o$.  Moreover,
it is shown in \dcite{k6015} that in the gauge \eqref{lf3} the
gauge potential $A^{\sigma} := \sigma^{*}\omega$ can be written
as
$A^{\sigma} = \tilde{A}^{\sigma} + \Lambda^{\sigma}
\hat{\sigma}^{*}\theta^{K}$
where $\tilde{A}^{\sigma} := \tilde{\sigma}^{*}\tilde{\omega}$,
$\Lambda^{\sigma} := \tilde{\Lambda}\circ \tilde{\sigma}$, and
$\theta^{K}_{h} := T_{h}\ell_{h^{-1}}$ ($\ell_h$ being the left
translation by $h$) is the Maurer-Cartan form on $K$.  This takes
care of the local formula for the gauge potential.

We now consider the Higgs field.  For $(y,k\Ko) \in \Ut\times\Uh$
we have $\sigma(y,k\Ko) =
[(\tilde{\sigma}(y),\hat{\sigma}(k\Ko)),e]$ and so it follows
from \eqref{HiggsLB} that $\Phi(y,k\Ko) =
[\sigma(y,k\Ko),\Lt\circ \tilde{\sigma}(y)]$ for all $(y,k\Ko)
\in \Ut\times\Uh$. This shows that in the gauge \eqref{lf3} (see
\eqref{PhiT}) the Higgs field is given by
\eqn{lf6}{
\phi^{\sigma}(y,k\Ko) = L^{\sigma}(y) := \Lt \circ
 \tilde{\sigma}(y) \quad (y,k\Ko) \in \Ut\times \Uh.
}
In view of \eqref{LmapsymA}, $L^{\sigma}$ must satisfy
$\rho_{\lambda(h)}L^{\sigma}(y) = L^{\sigma}(y) \quad \forall\;
h\in \Ko,\;\forall\; y\in \Ut.$
which becomes infinitesimally,
\leqn{lf9}{
\rho_{\lambda'(\xi)}L^{\sigma}(y) = 0 
\quad \forall\; \xi\in \kf_o, \;\forall\;y\in \Ut,
}
where $\rho : \go \longrightarrow \mathfrak{gl}(V)$ is the Lie
algebra representation of $\go$ on $V$ induced from the group
representation $\rho$.

Another important fact that will be needed for the analysis of
the EYMH equations is that invariant Higgs fields produce an
invariant stress-energy tensor. The Higgs field contribution to
the stress-energy tensor is made from the two following
combinations of the Higgs field:
$h(\Phi,\Phi) \quad \text{and} \quad h(D_{X}\Phi,D_{Y}\Phi)$
where $D_{X}$ is the covariant derivative on $E$ and $h$ is the
Hermitian metric on $E$ (see \eqref{Emetric}).

Equations \eqref{symloc}, \eqref{Emetric}, and the
$\Go$-invariance of $h$ imply that
\leqn{streng1}{
h(\psi^{E}_{k}\Psi,\psi^{E}_{k}\Phi) = h(\Psi,\Phi)
}
for all $(\Psi,\Phi) \in C\pi_{E} \times C\pi_{E}$ and $k\in
K$. Therefore any $K$-invariant section $\Phi$ satisfies
\leqn{streng2A}{
\psib_{k}^{*}\bigl(h(\Phi,\Phi)\bigr) = h(\Phi,\Phi) 
\quad \text{for all $k\in K$.}
}
This shows that $h(\Phi,\Phi)$ defines a $K$-invariant function
on $M$.

\sect{eqs}{Spherically symmetric field equations}

\subsect{lagreqs}{Field equations in general} 

We assume that the Lagrange density is $\Lagr \tau = \Lagr \sqrt{|g|}d^{4}x$
with $\Lagr$ given by \eqref{lagr}. Since for compact Lie groups all
finite-dimensional representations are equivalent to unitary ones the inner
product $h$ on $V$ can be assumed to be Hermitian and positive definite.  Here
$\DS\kk = \frac{c^{4}}{8\pi G}$ with $G$ being Newton's gravitation constant.
We may assume that any other physical coupling constants are subsumed in the
choice of the inner products. (For each simple component of a semisimple gauge
group $\Go$ and every irreducible subspace of $V$ there could be a different
coupling constant.) We will also use the notation $\norm{X}^{2}$ for $k(X,X)$
if $X\in\go$ and for $h(X,X)$ if $X\in V$.

Variation with respect to the metric, the gauge potential components and the
(real and imaginary) components of a Higgs field then yields the field
equations
\leqnarr{fee}{
\kk\bigl(R_{\alpha\beta}-\half R g_{\alpha\beta}\bigr) + \Lambda
g_{\alpha\beta} &=& T_{\alpha\beta} \label{fee1} \\
k(A,D^{\mu}F_{\mu\alpha})  &=&  2\,\Re\; h(\rho_{A}\Phi,D_\alpha\Phi) \quad
\forall\,A\in\go\label{fee2}\\
D^\mu D_\mu \Phi - 2\cW'\,\Phi &=& 0 \label{fee3}
}
where
\leqn{Tmn}{
\begin{split}
T_{\alpha\beta} = & k(F_{\alpha\mu},F_\beta^{\phantom{\beta}\mu}) 
            -\quart k(F_{\lambda\mu},F^{\lambda\mu}) g_{\alpha\beta} \\
  & + h(D_{(\alpha}\Phi,D_{\beta)}\Phi)
   - \half h(D_{\lambda}\Phi,D^{\lambda}\Phi) g_{\alpha\beta} 
   - \half \cW g_{\alpha\beta}.
\end{split}
}
Equation \eqref{fee2} can be written in the form
\leqn{fee1a}{
D^{\mu}F_{\mu\alpha} = \tilde{\rho}(\Phi,D_{\alpha}\Phi)
}
where $\tilde{\rho}: V \times V \ra \go$ is defined by 
\leqn{rhomap}{
\begin{split}
k(A,\tilde{\rho}(x,y)) &= h(\rho_Ax,y)+h(y,\rho_{A}x) \\
 &= h(y,\rho_{A}x) - h(x,\rho_{A}y)
       \fforall x,y\in V, \fforall A\in\go.
\end{split}
}
(the second formula being true because $\rho_{A}$ is an anti-Hermitian
operator on $V$).  It then follows from the invariance properties
of $k$ and
$h$,
\lgath{invp}{
k([A,B],C) = k(A,[B,C]) \quad\forall\; A,B,C \in\go, \label{invp1}\\
h(\rho_{A}x,y)+h(x,\rho_{A}y)=0 \quad\forall\;A\in \go,\,\forall x,y\in V, \label{invp2}
}
that the map $\tilde{\rho}$ satisfies
\leqnarr{rhorel}{
\tilde{\rho}(x,y) &=& -\tilde{\rho}(y,x)\, , \label{rhorel1} \\
k( [A,B],\tilde{\rho}(x,y) ) &=& k( A,\tilde{\rho}(\rho_Bx,y) )
- k( B,\tilde{\rho}(\rho_Ax,y) )\, . \label{rhorel2}
}
If
$k_{\Gamma\Delta}$ and $h_{IJ}$ are the components of $k$ and $h$ with
respect to bases $\{\eb_{\Gamma}\}$ of $\go$ and $\{\Eb_{I}\}$ of $V$,
respectively, then $\tilde{\rho}$ can be given by
\leqn{rhoup}{
\tilde{\rho}^{\Gamma}_{IJ} := - 2 k^{\Gamma\Sigma} \rho^K_{\Sigma [I}h_{J]K}
}
where $(k^{\Gamma\Delta})$ is the inverse matrix to $(k_{\Gamma\Delta})$ and
$\rho_{\eb_\Gamma}(\Eb_{J}) = \Eb_{K}\rho^K_{\Gamma J}$.

In the special case where $\rho$ is
the adjoint representation, the map $\tilde{\rho}$ is given
by the negative of the Lie bracket, i.e. $\tilde{\rho}(A,B) = -[A,B]$.

\subsect{sph}{Spherically symmetric EYMH fields}

The spherically symmetric space-time metric can be given in a
Schwarzschild-like coordinate system by
\leqn{statmetric}{
g = -N S^{2}\, dt^{2}+
  N^{-1}\,dr^{2}+r^{2}\bigl(d\theta^{2}+\sin^{2}\theta \,d\varphi^{2}\bigr).
}
where $N$ and $S$ are functions of $r$ and $t$, in general, and of $r$ only in
the static case. The function $N$ is related to the mass function $m(r,t)$ by
$N=1-2m/r-\Lambda/(3\kk)r^{2}$. We assume that the space-time $M$ is
diffeomorphic to $\Mt\times S^{2}$ where $\Mt$ is the `r-t' manifold and
$S^{2}$ the orbits of the symmetry group action.

The Yang-Mills potential for the gauge group $G=SU(2)$ has often been given in
the so-called Witten form \dcite{k4864} which is, however, not easily
generalized to other gauge groups. Potentials for general compact gauge groups
(in the EYM case) have first been discussed by Bartnik \dcite{k5109} and by
Brodbeck and Straumann \dcite{k5281}. They show that the gauge potential can
be given in the form
\leqn{statconnx}{
A = NS\Ac dt + \Bc dr + \La_{1} d\theta + 
    \big(\La_{2}\sin\theta + \La_{3}\cos\theta\bigr)d\varphi.
}
If we choose the symmetry group to be $K=SU(2)$ whose action on space-time has
as isotropy subgroup $\Ko=U(1)$ so that $K/\Ko\simeq S^{2}$ then $\La$ is a
map from $\Mt$ into the space of linear maps from $\kf$ to $\go$ subject to
\eqref{CCthm2} which implies
\leqn{wang}{%
 [\La_{2},\La_{3}]=\La_{1} \AND [\La_{3},\La_{1}]=\La_{2},
}
where $\La_{k}=\La(\tau_{k})$, $\{\tau_{k}\,:\,k=1,2,3\}$ being the standard basis
of $\su(2)$ with $\tau_{3}$ spanning $\kf_{o}$.  So
$\La_{3}=\lambda'(\tau_{3}) \in \go$ is a constant vector characterizing
the embedding of $SU(2)$ in $G$ and thus the conjugacy class of the
$SU(2)$-action on $P$.  Also $\Ac$ and $\Bc$ are $\go$-valued functions on
$\Mt$ which, moreover, commute with $\La_{3}$.  They give the ``electric''
part of the Yang-Mills potential. One can choose a temporal gauge so that
$\Bc=0$, and since one is mostly interested in the noncommuting aspects of the
Yang-Mills field the component $\Ac$ is often assumed to be zero, as we will
also do from now on.

The static spherically symmetric field equations for the full EYMH system can
now be written in a form just slightly more general than those derived in
\dcite{eymg,eymgg}. We need to observe that locally invariant Higgs fields are
described by $V$-valued functions of $r$, i.e maps $r\in \tilde{U}\subset
\Rset \ra V$, since here $H=U(1)$, subject to the condition \eqref{lf9} which
becomes 
\leqn{hcond}{
\rho_{\La_{3}} \Phi(r) = 0.
}
The Yang-Mills equations then become
\leqnarr{sfeq}{
r^{2}S^{-1}(NS\Lambda'_{1})' - [\Lambda_{2},\Fh] &=& r^{2}
\tilde{\rho}(\Phi,\rho_{\Lambda_{1}}\Phi) \label{sfeq1} 
\\
r^{2}S^{-1}(NS\Lambda'_{2})' + [\Lambda_{1},\Fh] &=& r^{2}
\tilde{\rho}(\Phi,\rho_{\Lambda_{2}}\Phi) \label{sfeq2} 
\\
{[}\Lambda'_{1},\Lambda_{1}] + [\Lambda'_{2},\Lambda_{2}] &=& r^{2}
\tilde{\rho}(\Phi,\Phi') \label{sfeq3}
}
where
\leqn{Fdef}{
\Fh := [\Lambda_{1},\Lambda_{2}] - \Lambda_{3}.
}
The Higgs equation takes the form
\leqn{higgseq}{
S^{-1}( r^{2} N S \Phi' )' + 
 ( \rho_{\Lambda_{1}}\rho_{\Lambda_{1}} + \rho_{\Lambda_{2}}\rho_{\Lambda_{2}}
 ) \Phi + \cW'\, \Phi = 0.
}
To derive the expression for the stress-energy tensor repeated use of 
\eqref{wang}, \eqref{hcond}, the representation property,
$\rho_{[X,Y]}=\rho_{X}\rho_{Y}-\rho_{Y}\rho_{X}$ for $X,Y\in\go$, as well as
the assumption that $\rho_{X}$ is anti-Hermitian on $V$ must be made. 
We find from these relations that
\lgath{rels}{
[\La_{3},\Fh] = 0,\quad k(\Lambda'_{1},\Fh) = k(\Lambda'_{2},\Fh) = 0, \label{rels1}\\
k(\Lambda_{1},\Lambda_{2}) = k(\Lambda'_{1},\Lambda'_{2}) = 
k(\Lambda_{2},\Lambda_{2}) - k(\Lambda_{1},\Lambda_{1}) =
k(\Lambda'_{2},\Lambda'_{2}) - k(\Lambda'_{1},\Lambda'_{1}) = 0, \label{rels2}\\
h(\rho_{\Lambda_{2}}\Phi,\rho_{\Lambda_{2}}\Phi) =
h(\rho_{\Lambda_{1}}\Phi,\rho_{\Lambda_{1}}\Phi), \quad
h(\rho_{\Lambda_{2}}\Phi,\rho_{\Lambda_{1}}\Phi) =
-h(\rho_{\Lambda_{1}}\Phi,\rho_{\Lambda_{2}}\Phi) \label{rels5}
} 
and then that $(T^{\alpha}_{\beta}) = \diag( -e, p_{r}, p_{\theta}, p_{\theta})$ with
\lalign{stress}{
e          &= r^{-2}N \norm{\Lambda'_{1}}^{2} &+&\half r^{-4}\norm{\Fh}^{2} &+&\half N\norm{\Phi'}^{2} &+& r^{-2} \norm{\rho_{\Lambda_{1}}\Phi}^{2} &+&\half \cW,\label{stress1} \\
p_{r}      &= r^{-2}N \norm{\Lambda'_{1}}^{2} &-&\half r^{-4}\norm{\Fh}^{2} &+&\half N\norm{\Phi'}^{2} &-& r^{-2} \norm{\rho_{\Lambda_{1}}\Phi}^{2} &-&\half \cW,\label{stress2} \\
p_{\theta} &=                       &\phantom{-}&\half r^{-4}\norm{\Fh}^{2} &-&\half N\norm{\Phi'}^{2} & &                                          &-&\half \cW.
\label{stress3}
}
so that the Einstein equations become
\lalign{einstein}{
\kk\, m'  &= \half r^{2} e,\label{einstein1}\\
\kk\, S^{-1}S' &= \half r N^{-1} ( e + p_{r} ) = r^{-1}
\norm{\Lambda'_{1}}^{2} + \half r \norm{\Phi'}^{2}. \label{einstein2}
} 

\subsect{cons}{Consistency of the spherically symmetric equations} 

The equation \eqref{sfeq3} can be viewed as a constraint equation since the
equations \eqref{sfeq1}, \eqref{sfeq2}, and \eqref{higgseq} are second order
differential equations which when solved will fully determine Yang-Mills
potential and the Higgs field.  The next proposition shows that away from the
singular points where $N(r)=0$, $S(r)=0$, or $r=0$ the constraint equation
\eqref{sfeq3} is `conserved', i.e. automatically satisfied if it is satisfied
at one point and hence it is only a constraint on the initial data for the
differential equations \eqref{sfeq1},\eqref{sfeq2}, \eqref{higgseq}. We
suspect that as in the EYM case this will still hold for solutions defined
about the singular point but we have not (yet) done an analysis of the
differential equation near the singular points similar to that in
\dcite{eymg,eymgg}.

\begin{prop} \label{consp}
  Suppose $\{N(r),S(r),\Lg_1(r),\Lg_2(r)\}$ satisfy the Yang-Mills equations
  \eqref{sfeq1} and \eqref{sfeq2} and the Higgs equation \eqref{higgseq} 
  on an interval $[r_1,r_2)$ $(r_1>0)$.  If neither $N(r)$ nor
  $S(r)$ vanish on the interval $[r_1,r_2)$ and if the constraint equation 
  \eqref{sfeq3} holds at $r=r_{1}$ then it holds at all $r\in[r_{1},r_{2})$.
\end{prop}

\begin{pf}
Let
\eqn{conc3}{
\gamma := [NS\Lg_1',\Lg_1]+[NS\Lg_2',\Lg_2]
- \rhoi(\Phi,r^2 NS\Phi'). }
Differentiating $\gamma$ and using equations
\eqref{sfeq1}, \eqref{sfeq2}, and \eqref{higgseq}
yields
\lalign{conc4}{
\gamma' = r^{-2}S\bigl(&[[\Lg_2,\Fh],\Lg_1]
-[[\Lg_1,\Fh],\Lg_2]\bigr)\notag\\ &+ S\bigl(
[\rhoi(\Phi,\rho_{\Lg_1}\Phi),\Lg_1]
+[\rhoi(\Phi,\rho_{\Lg_2}\Phi),\Lg_2]
+ \rhoi(\Phi,(\rho^2_{\Lg_1}+\rho^2_{\Lg_2})\Phi)
\bigr)\, . \label{conc4.1}
}
The Jacobi identity and equation \eqref{wang} imply that
\leqn{conc5}{
[[\Lg_2,\Fh],\Lg_1] -[[\Lg_1,\Fh],\Lg_2] = 0\, ,
}
while for any $A\in \go$ and $j=1,2$,
\leqn{conc6}{
k(A,[\rhoi(\Phi,\rho_{\Lambda_j}\Phi),\Lambda_j])
= -k(A,\rhoi(\Phi,\rho_{\Lambda_j}^{2}\Phi))
+k(\Lambda_j,\rhoi(\Phi,\rho_A\rho_{\Lambda_j}\Phi)) 
}
follows from \eqref{invp1}, \eqref{rhorel1}, and \eqref{rhorel2}.
But
\alin{conc7}{
k(\Lambda_j,\rhoi(\Phi,\rho_A\rho_{\Lambda_j}\Phi)) &
\stackrel{\eqref{rhomap}}{=}
h(\rho_{\Lambda_j}\Phi,\rho_{A}\rho_{\Lambda_j}\Phi) +
h(\rho_A\rho_{\Lambda_j}\Phi,\rho_{\Lambda_j}\Phi) 
\\
& \stackrel{\phantom{\eqref{rhomap}}}=
-h(\rho_A\rho_{\Lambda_j}\Phi,\rho_{\Lambda_j}\Phi) +
h(\rho_A\rho_{\Lambda_j}\Phi,\rho_{\Lambda_j}\Phi) = 0
}
since $\rho_A$ is anti-Hermitian. Since $A\in \go$ was chosen arbitrarily,
\eqref{conc6} then implies that
\leqn{conc8}{
[\rhoi(\Phi,\rho_{\Lg_1}\Phi),\Lg_1]
+ [\rhoi(\Phi,\rho_{\Lg_2}\Phi),\Lg_2]
+ \rhoi(\Phi,(\rho^2_{\Lg_1}+\rho^2_{\Lg_2})\Phi)
= 0 \, .
}
So \eqref{conc4.1}, \eqref{conc5} and \eqref{conc8} imply that $\gamma'=0$
and hence $\gamma = \text{const}$ on the interval $[r_1,r_2)$. Clearly
$\gamma(r_1)=0$ then implies that $\gamma(r)=0$ for all $r\in [r_1,r_2)$.
\end{pf}

It remains to investigate the consistency of the Yang-Mills equations
\eqref{sfeq1} and \eqref{sfeq2} together with \eqref{wang}. First we have

\begin{prop} \label{Lcom}
Let $\Rt_{j} := \rhoi(\Phi,\rho_{\Lambda_j}\Phi)$ for
$j=1,2$. Then 
\leqn{Lcom1}{
[\Rt_2,\Lg_3] = \Rt_1 \quad \text{and} \quad [\Lg_3,\Rt_1] = \Rt_2 \, .
}
\end{prop}

\begin{pf}
Suppose $A\in \go$. Then
\alin{Lcom2}{
k(A,[\Lg_3,\Rt_1]) &\stackrel{\phantom{\eqref{rhorel2}}}{=} k(A,[\Lg_3,
\rhoi(\Phi,\rho_{\Lg_1}\Phi)]) \stackrel{\eqref{invp1},\eqref{rhorel1}}{=}
-k([A,\Lg_3],\rhoi(\rho_{\Lg_1}\Phi,\Phi)) &&
\\
& \stackrel{\eqref{rhorel2}}{=} -k(A,\rhoi(\rho_{\Lg_3}\rho_{\Lg_1}\Phi,\Phi))
+k(\Lg_3,\rhoi(\rho_A\rho_{\Lg_1}\Phi,\Phi)) &&
\\ 
& \stackrel{\eqref{wang}}{=}
-k(A,\rhoi((\rho_{\Lg_2}+\rho_{\Lg_1}\rho_{\Lg_3})\Phi,\Phi))
+k(\Lg_3,\rhoi(\rho_A\rho_{\Lg_1}\Phi,\Phi)) &&
\\
& \stackrel{\eqref{hcond}}{=} k(A,\rhoi(\Phi,\rho_{\Lg_2}\Phi))
+k(\Lg_3,\rhoi(\rho_A\rho_{\Lg_1}\Phi,\Phi)) && 
\\
&\stackrel{\phantom{\eqref{rhorel2}}}{=} k(A,\Rt_2) +
k(\Lg_3,\rhoi(\rho_A\rho_{\Lg_1}\Phi,\Phi))\, .
}
But
\alin{Lcom3}{
 k(\Lg_3,\rhoi(\rho_A\rho_{\Lg_1}\Phi,\Phi))
& \stackrel{\eqref{rhomap}}{=} h(\rho_{\Lg_3}\rho_A\rho_{\Lg_1}\Phi,\Phi)
+ h(\Phi, \rho_{\Lg_3}\rho_A\rho_{\Lg_1}\Phi)
\\
& \stackrel{\phantom{\eqref{rhomap}}}{=} -
h(\rho_A\rho_{\Lg_1}\Phi,\rho_{\Lg_3}\Phi) -
h(\rho_{\Lg_3}\Phi,\rho_A\rho_{\Lg_1}\Phi) 
\\
& \quad\text{\ \ (since $\rho_{\Lg_3}$ is anti-Hermitian)} 
\\
& \stackrel{\eqref{hcond}}{=} 0\,. 
}
Since $A\in \go$ was chosen arbitrarily, the above two
results imply that $ [\Lg_3,\Rt_1]=\Rt_2$. Similar calculations
show that $[\Rt_2,\Lg_3] = \Rt_1$. 
\end{pf} 
With \eqref{Lcom1} it follows easily that \eqref{sfeq1} and \eqref{wang}
together imply \eqref{sfeq2}.
In fact, these Yang-Mills equations are more conveniently described in
complex form.  Let $\g=\go\otimes\Cset$ be the complexification of $\go$ so
that $\go$ is its compact real form with respect to the conjugation
$c:\g\ra\g:X+iY \mapsto X-iY\;\forall \;X,Y\in\go$ and let
\leqn{lamdef}{
\Lo = 2i\La_{3},\quad \Lpm := \mp\La_{1}-i\La_{2}
} 
so that $\Lm = -c(\Lp)$ and $c(\Lo)=-\Lo$ and, by \eqref{wang},
\leqn{wangc}{
[\Lo,\Lpm] = \pm 2 \Lpm\,.
}
The Yang-Mills equations \eqref{sfeq1},\eqref{sfeq2} are then equivalent to
\leqn{ymeq}{
r^{2}S^{-1}(NS\Lp')' - i[\Fh,\Lp] = - r^{2}(\Rt_{1} + i \Rt_{2}) =:
-r^{2}\Rt_{+}\,.
}
With respect to the invariant metric $k$ the operator $\ad_{\Lo}$ is
Hermitian and $\g$ can be decomposed into eigenspaces of $\ad_{\Lo}$,
\leqn{esp0}{
\g = \bigoplus \g_{n}, \quad \g_{n} := \{ X\in\g \,:\, [\Lo,X]=n X \}\, .
}
By \eqref{wangc}, $\Lp(r) \in \g_{2} \,\forall\, r$ and therefore so are
$\Lp'$, $\Lp''$, and also, by \eqref{conc5}, $[\Fh,\Lp]$. On the other
hand, Proposition \ref{Lcom} implies that also the right hand side of
\eqref{ymeq} lies in $\g_{2}$ for any (anti-Hermitian) representation
$\rho:\go \ra \gl(V)$ provided that the Higgs field satisfies
\eqref{hcond}. Equations \eqref{ymeq} thus represent consistent second
order differential equations for the gauge potential components, subject
only to the constraints \eqref{sfeq3} being satisfied at one point.

\subsect{bdc}{Explicit form of the field equations}

If we are just trying to construct a local solution of the EYMH equations in
some radial interval in which none of $r$, $N(r)$ and $S(r)$ is zero we can
choose a constant $\Lo \in \g$, subject to it being an integral lattice point
within the closed fundamental Weyl chamber of some Cartan subalgebra and
satisfying $c(\Lo)=-\Lo$. (This will fix an explicit action of the symmetry
group $K_{o}=SU(2)$ by automorphisms on the principal bundle \dcite{k5281}.) Then
any $r$-dependent $\Lp \in\g$ may be chosen subject to \eqref{wangc} and, at
one point, to \eqref{sfeq3}.

But the interesting and physically more relevant EYMH fields are global ones
which remain regular at the center $r=0$ or at a black hole horizon where
$N=0$ and which have an appropriate asymptotic behavior. It is clear from the
expressions for energy density and pressures in equations
\eqref{stress1}-\eqref{stress3} that $\Lambda'_{1}$, $\rho_{\Lambda_{1}}\Phi$
and, in particular, $\Fh$ must vanish for $r=0$.  This means that also
$[\La_{1},\La_{2}]=\La_{3}$ at that point which in turn implies that the
induced Lie algebra homomorphism $\lambda': \mathfrak{\kf_{0}}\ra\go$ defines
a so-called $A_{1}$ (or defining) vector $\Lo = 2i\,\Lambda_{3}$ in the Cartan
subalgebra of the complexified Lie algebra $\g=\go\otimes\Cset$ and thus a
conjugacy class of $\sL(2)$-subalgebras.  Even when no regularity at the
center is required, for example when solutions need only be found outside a
black hole, natural physical fall-off conditions at infinity also imply
$\Fh=0$ (at least when the space-time is asymptotically flat and the magnetic
charge vanishes). We will therefore from now on make the assumption that $\Lo$
is an $A_{1}$-vector.

Up to conjugacy these $A_{1}$-vectors and their corresponding subalgebras form
a finite set and, given a base $\{\alpha_{1},\ldots,\alpha_{\ell}\}$ of the
set of roots $R$ of the Lie algebra $\g$, are uniquely described by the
characteristic $\chi = \bigl(\alpha_{1}(\Lo), \ldots,
\alpha_{\ell}(\Lo)\bigr)$.
There is always a root base $\Delta$ such that
$\alpha_{k}(\Lo)\in \{0,1,2\}\,\forall\,k$, and all possible characteristics
and thus all conjugacy classes of $\sL(2))$-subalgebras of simple Lie algebras
have been classified (\dcite{k4779,k6494}).  In view of \eqref{wangc} the
invariant connection on the principal bundle for a given conjugacy class of
$K$-actions of automorphisms is then fully given by the (complex) functions
$w_{\alpha}(r)$ such that
\leqn{lam}{
\Lp=\sum_{\alpha\in \Sset} w_{\alpha} \eb_{\alpha},\quad 
\Sset := \{ \alpha\in R \,:\, \alpha(\Lo)=2\ \}\,.
}
Here we have introduced a Chevalley-Weyl basis $\{\hb_{\alpha},\eb_{\beta},
\eb_{-\beta} \,:\, \alpha\in \Delta,\beta\in R^{+}\}$ (where $R^{+}$ is the
set of positive roots, cf., for example, \dcite{k5157}) of $\g$ for which
we adopt the conventions and definitions
\footnote{
If the gauge group is semisimple and an invariant inner product on $\g$
contains more than one `coupling' constant this may have to be modified.
}
\lgath{CWbas}{
 [\eb_{\alpha},\eb_{-\alpha}] = \hb_{\alpha},\quad
 [\eb_{\alpha},\eb_{\beta}] = \nu_{\alpha,\beta}\, \eb_{\alpha+\beta},\quad
 \nu_{-\alpha,-\beta}=-\nu_{\alpha,\beta} \text{\ or\ } 0,\text{\ if\ }
 \alpha+\beta\notin R,
\\
\abs{\alpha}^{2} := k(\alpha,\alpha), \quad
\langle \alpha,\beta \rangle :=
\frac{2k(\alpha,\beta)}{\abs{\beta}^{2}} \quad \forall\,\alpha,\beta\in R,
\label{CWbas2}
\\
 k(\eb_{\alpha},\eb_{-\alpha}) = -2 \abs{\alpha}^{-2} \quad
 \forall\,\alpha,\beta\in
 R, \quad
(c_{ij}) := \bigl( \langle \alpha_{i}, \alpha_{j} \rangle \bigr)\quad
 \text{(Cartan matrix)}\,. \label{CWbas3}
}

The gauge connection is thus described by as many complex functions of $r$
as there are elements in $\Sset$. In fact, by the definition of the roots,
the eigenspace $\g_{2}$ of $\ad_{\Lo}$ is spanned by the set $\{
\eb_{\alpha} \,:\, \alpha\in \Sset \}$.

In the EYM case the field equations need to be solved for the two real
functions $N$ (or $m$) and $S$ of $r$ and the complex functions
$w_{\alpha}=\omega_{\alpha}e^{i \gamma_{\alpha}} = u_{\alpha} + i
v_{\alpha}$ for $\alpha\in \Sset$. This turns out to be considerably
simpler if the set $\Sset$ forms a $\Pi$-system \dcite{k4779}, i.e. if
$\alpha,\beta\in \Sset$ implies that $\alpha-\beta$ is not a root. Then
$[\eb_{\alpha}, \eb_{-\beta}]=0$ if $\alpha$ and $\beta$ are two distinct
elements of $\Sset$ \dcite{k5281,eymg}. Then $\Sset$ also generates a
subalgebra of $\g$. In particular, $\Sset$ is a $\Pi$-system if $\Lo$ is
contained in the open Weyl chamber of the Cartan subalgebra of $\g$
\dcite{k5281} which means, in particular, that
$\alpha(\Lo)>0\;\forall\;\alpha\in R^{+}$. We have called this the
\emph{regular case}.

The simplification occurs largely because the constraint equation
\eqref{sfeq3} then implies that the phase $\gamma_{\alpha}$ of $w_{\alpha}$
is constant and can be chosen zero by a gauge choice. As the following
shows this may not always be the case in the presence of Higgs fields, but
the equations are still much simpler.

In the following we will derive an explicit form for the Yang-Mills and the
Higgs equations only since no new insight is gained by reformulating
Einstein's equations.

The left hand side of equations \eqref{sfeq1}-\eqref{sfeq3} has been derived
in \dcite{eymg} and \dcite{eymgg}. From \eqref{lam} 
we have
\leqn{ymexp}{
\begin{split}
r^{2} S^{-1}(NS w'_{\alpha})' &+ \half \biggl( \alpha(\Lo) w_{\alpha} -
\DS\sum_{\beta\in\Sset} \langle \beta,\alpha \rangle
\abs{w_{\alpha}}^{2}w_{\beta} + \DS \sum_{\beta,\gamma,\delta \in \Sset}
\mu_{\alpha\delta\beta\gamma}\, w_{\beta} \bar{w}_{\gamma}w_{\delta} \bigg)
\\
& = -r^{2} \Rt_{+,\alpha} \;\;\forall\;\alpha\in \Sset
\end{split}
}
and
\leqn{ymcstr}{
\sum_{\alpha,\beta \in \Sset} ( w_{\alpha}\bar{w}'_{\beta} -
w'_{\alpha}\bar{w}_{\beta} ) [ \eb_{\alpha},\eb_{-\beta} ] = 2 r^{2}
\rhoi(\Phi,\Phi')
}
where
\[
[\eb_{\alpha},[\eb_{\beta},\eb_{-\gamma}]] =: \sum_{\delta \in \Sset}
\mu_{\delta\alpha\beta\gamma} \eb_{\delta}
\]
and $\Rt_{+,\alpha}$ is the $\eb_{\alpha}$-component of $\Rt_{+}$. Note that
it follows from proposition \ref{Lcom} that $\Rt_{+} \in \g_{2} = \spann
\{\eb_{\alpha}\,:\,\alpha\in \Sset \}$. Moreover, $[\eb_{\alpha},
\eb_{-\beta}] \in \g_{0}$ if $\alpha,\beta \in \Sset$ and $\rhoi(\Phi,\Phi')
\in \g_{0}$. By proposition \ref{consp} \eqref{ymcstr} needs to be solved for
the $w'_{\alpha}$'s only for one $r$-value.

In the \emph{regular case} $\mu_{\alpha\beta\gamma\delta}=0$, so
\eqref{ymexp} represents the components of an equation in the span of
$\{\hb_{\alpha} : \alpha\in \Sset\}$. Moreover, $[\eb_{\alpha},
\eb_{-\beta}] = \delta_{\alpha\beta}\hb_{\alpha}$ so that \eqref{ymcstr}
becomes a condition for the derivatives of the phases of the complex
functions $w_{\alpha}(r)$ (which when the right hand side vanishes like in
the EYM case means that the phases will be constant and the $w_{\alpha}$s
can be chosen real by fixing the gauge.)

Since in \eqref{rhomap} the quantity $\rhoi$ is only defined for $A\in \go$
in order to evaluate the right hand side of \eqref{ymexp} and \eqref{ymcstr}
we introduce (temporarily) the basis
\leqn{rbase}{
\hat{\hb}_{j} := -\ihalf \hb_{j},\; 
\hat{\eb}_{\alpha} := \half(-\eb_{\alpha}+\eb_{-\alpha}),\;
\hat{\fb}_{\alpha} := \ihalf(\eb_{\alpha}+ \eb_{-\alpha}) \quad
(j=1,\ldots \ell,\; \alpha\in R^{+}).
}
whose $\Rset$-linear span is the compact real form $\go$ of $\g$. In this
basis $\{ \eb_{\Gamma} \} = \{ \hb_{j}, \hat{\eb}_{\alpha},\hat{\fb}_{\alpha}
\}$ the invariant metric then has the form
\leqn{rkmat}{
\bigl( \hat{k}_{\Gamma\Delta} \bigr) = \begin{pmatrix}
\half \abs{\alpha_{i}}^{-2} c_{ij} & 0 & 0 \\
0 & \abs{\alpha}^{-2} \delta_{\alpha\beta} & 0 \\
0 & 0 & \abs{\alpha}^{-2} \delta_{\alpha\beta}
\end{pmatrix}
}

We extend the anti-Hermitian representation $\rho:\go \ra \gl(V)$ to $\g$ in
the obvious way, $\rho_{X+iY}:=\rho_{X}+i\,\rho_{Y}$, and let
\leqn{rhodef}{
\rho_{j} := \rho_{\hb_{j}}\,\forall j=1\ldots\ell 
\quad\text{and}\quad
\rho_{\alpha} := \rho_{\eb_{\alpha}} \,\forall\,\alpha\in R^{+}\,.
}
It then follows for the Hermitian conjugates with respect to the inner product
$h$ on $V$ that
\leqn{herm}{
\rho^{+}_{j} = \rho_{j},\quad \rho^{+}_{\alpha} = \rho_{-\alpha}
}
and that $\rho_{\Lo}$ is Hermitian and $\rho_{\Lpm}^{+}=\rho_{\Lmp}$. Denoting
the inverse of $k(\hat{\hb}_{i},\hat{\hb}_{j})$ by
\leqn{kinv}{
\hat{k}^{ij} = 2 c^{ij}\abs{\alpha_{j}}^{2}\,,
}
where $(c^{ij})$ is the inverse of the Cartan matrix, we find from
\eqref{rhomap} that

\leqn{rhot}{
\begin{split}
\tilde{\rho}(x,y) = -\sum_{i,j=1}^{\ell} \hat{\hb}_{i}\,\hat{k}^{ij}\, \Im\;
h(\rho_{j}x,y) - \sum_{\alpha\in R^{+}} & \abs{\alpha}^{2} \bigg( 
\Re \bigl[h(\rho_{\alpha}x,y) - h(x,\rho_{\alpha}y) \bigr]\, \hat{\eb}_{\alpha} 
\\
& -\Im \bigl[ h(\rho_{\alpha}x,y) + h(x,\rho_{\alpha}y) \bigr]\,
\hat{\fb}_{\alpha} \biggr)\,.
\end{split}
}

Now any (finite-dimensional) representation of $G$ is the direct sum of
irreducible ones which can be obtained from irreps of $\g$ and are
characterized by their highest weight $\Lambda \in \h^{*}$ (the dual of the
Cartan subalgebra $\h$).  Any other weight $\mu$ is then given by $\mu =
\Lambda - \sum_{i=1}^{\ell} q_{i} \alpha_{i}$ for certain nonnegative integers
$q_{i}$.  The set of eigenvalues of $\rho_{\Lo}$ is $\mathcal{E}_{o}=\{
\mu_{k}(\Lo) \}$, where the $\mu_{k}$ are the weights of the representation.
Thus spherically symmetric Higgs fields for a given $\Lo$, i.e. choice of the
action of $K$, and a given representation $\rho$ exist provided at least one
of the irreducible components of $\rho$ has a weight $\mu$ with $\mu(\Lo) =
0$.

In particular, for the adjoint representation, $\rho=\ad$ and $V=\go$, there
is always a weight $0$ with multiplicity equal to the rank of $\g$ and any
$\Phi \in \h$ is a solution of \eqref{hcond}. 

Moreover, in the \emph{regular case} where $\alpha(\Lo)>0$ for all positive roots
$\alpha$ every solution of \eqref{hcond} lies in $\h$.

Next we observe that since $\rho_{\Lo}$ is a Hermitian operator the vector
space $V$ is a direct sum of mutually orthogonal eigenspaces of $\rho_{\Lo}$,
\leqn{Vdec}{
V = \bigoplus_{\sigma\in\mathcal{E}_{o}} \hat{V}_{\sigma},\quad
\hat{V}_{\sigma} := \{ x \in V \;:\; \rho_{\Lo}x = \sigma x \}
}
so that \eqref{hcond} now states that $\Phi(r) \in \hat{V}_{0} \;\forall\;r$
and thus also $\Phi'(r)\in \hat{V}_{0}$.  Moreover, it follows easily that
\leqn{rV}{
\rho_{\alpha} V_{\sigma} \subset V_{\sigma+\alpha(\Lo)} \quad\text{and}\quad 
\rho_{h} V_{\sigma} \subset V_{\sigma}\;\forall\;h\in \h
}
and therefore $\rho_{\alpha}\Phi \in \hat{V}_{2}$ when $\alpha \in \Sset$
so that, in particular,
\leqn{rpmPhi}{
\rho_{\Lpm}\Phi \in \hat{V}_{\pm2} \quad\text{and}\quad \rho_{j}\Phi \in \hat{V}_{0}\,.
}
Thus, if we replace $x$ and $y$ in \eqref{rhot} by $\Phi$ and $\Phi'$,
respectively, we get the expression needed on the right hand side of
\eqref{ymcstr} except that the second sum needs only be taken over those
roots $\alpha\in R^{+}$ for which $\alpha(\Lo)=0$, in view of \eqref{rV},
since both $\Phi$ and $\Phi'$ lie in $\hat{V}_{0}$ and the
$\hat{V}_{\sigma}$ are orthogonal for distinct $\sigma$.

In the \emph{regular case} $\alpha(\Lo)>0\;\forall\;\alpha\in\Sset$ so that
the constraint equation \eqref{ymcstr} becomes
\leqn{ymcstrr}{
\sum_{j=1}^{\ell} \hat{k}_{ij} a_{j}\abs{\alpha_{j}}^{2}
\abs{\alpha}^{-2} \omega_{\alpha}^{2}\,\gamma'_{\alpha}  = -\half r^{2} \Im \,
h(\rho_{i}\Phi,\Phi')\quad\forall\, i=1\ldots \ell, \; \forall\, \alpha\in
\Sset
}
where $\alpha = \sum_{j=1}^{\ell} a_{j}\alpha_{j}$.

In the evaluation of $\Rt_{+}$ one obtains expressions
$h(\rho_{\pm\alpha}\Phi,\rho_{\pm\beta}\Phi)$ (for all choices of the signs)
where $\alpha\in \Sset$ and $\beta\in R^{+}$. But if $\Phi\in \hat{V}_{0}$
then $\rho_{\pm\alpha}\Phi\in \hat{V}_{\pm2}$ and $\rho_{\pm\beta}\Phi \in
\hat{V}_{\pm\beta(\Lo)}$. Since these eigenspaces of $\rho_{\Lo}$ are mutually
orthogonal the only inner products that are nonzero are those when
$\alpha,\beta \in \Sset$ and the signs are the same. It follows that
\leqn{Rp}{
\Rt_{+} = \sum_{\alpha \in \Sset} \Rt_{+,\alpha} \eb_{\alpha} =
\sum_{\alpha \in \Sset} \abs{\alpha}^{2} Q_{\alpha} \eb_{\alpha}
}
where
\leqn{Qdef}{
Q_{\alpha} := \half \DS \sum_{\beta \in \Sset} \bigl( R_{\alpha,\beta} +
\overline{R}_{-\alpha,-\beta} \bigr) w_{\beta}
}
and
\leqn{Rdef}{
R_{\alpha,\beta} := h(\rho_{\alpha}\Phi,\rho_{\beta}\Phi),\quad
\alpha,\beta \in R,\; \Phi \in \hat{V}_{0} \,.
}

Again, in the \emph{regular case}, or whenever we know that $\alpha,\beta
\in \Sset$ implies that $\alpha-\beta$ is not a root, this simplies
somewhat. For we have
$\DS \rho_{-\alpha}\rho_{\beta}\Phi =
\rho_{[\eb_{-\alpha},\eb_{\beta}]}\Phi + \rho_{\beta}\rho_{-\alpha}\Phi =
\rho_{\beta}\rho_{-\alpha}\Phi$ 
and therefore
$R_{-\alpha,-\beta} = h(\rho_{-\alpha}\Phi, \rho_{-\beta}\Phi) =
h(\Phi,\rho_{\alpha}\rho_{-\beta}\Phi) =
h(\Phi,\rho_{-\beta}\rho_{\alpha}\Phi) = h(\rho_{\beta}\Phi,
\rho_{\alpha}\Phi) = \overline{ h(\rho_{\alpha}\Phi, \rho_{\beta}\Phi)}
= \overline{R}_{\alpha,\beta}$
so that
\leqn{Qr}{
Q_{\alpha} := \sum_{\beta \in \Sset} R_{\alpha,\beta}w_{\beta} \,.
}

\sect{concl}{Conclusions}

We have shown how to, in principle, construct Einstein-Yang-Mills-Higgs
systems that are invariant under an arbitrary action of a space-time symmetry
group that acts by principle bundle automorphisms which leave the gauge
connection invariant as well as Higgs fields defined via any unitary
representation of the (compact) gauge group.  The classification of the
possible actions by automorphisms is known for the symmetry group $SU(2)$, but
may be more difficult to find for larger groups. One would need to first find
all conjugacy classes of a certain type of Lie subalgebras of the gauge Lie
algebra.

We have obtained an explicit form of the full field equations in the static
spherically symmetric case and shown that they form a consistent system of
ordinary differential equations. Before global solutions can be found
numerically it would be necessary to investigate in some detail the boundary
conditions that regularity conditions at a center, horizon or in an asymptotic
region will imply.

It must be pointed out that not all cases of physical interest even for the
static spherically symmetric case are covered by this approach. For example,
the doublet Higgs field coupled to an $SU(2)$-gauge and gravitational field in
\dcite{k5569} cannot be described in our formalism because the Higgs field is
not spherically symmetric. In fact, the representation of $\su(2)$ in this
case is the direct sum of two irreducible two-dimensional ones for which there
is no weight $\mu$ with $\mu(\Lo)=0$. These authors make a simple ansatz for
the Higgs field using the gauge choice for the potential often attributed to
Witten \dcite{k4864}. They then find that the stress-energy tensor is
spherically symmetric and thus compatible with a spherically symmetric ansatz
for the space-time metric.

One might ask whether with our gauge choice one can assume that only the
quantity $h(D_{(\alpha}\Phi,D_{\beta)}\Phi)$ is spherically symmetric rather
than $\Phi$ itself. Unfortunately this is not possible since then
$h(\rho_{\Lambda_{3}}\Phi,\rho_{\Lambda_{3}}\Phi)$ would have to vanish which
implies $\rho_{\Lambda_{3}}\Phi=0$, i.e. an invariant Higgs field.  On the
other hand we do not know whether the Witten ansatz for spherically symmetric
gauge fields can be generalized to gauge groups other than $SU(2)$ or whether
perhaps another equally convenient gauge exists.


\end{document}